# BRONCO: Automated modelling of the bronchovascular bundle using the Computed Tomography Images

Wojciech Prażuch[1§], Marek Socha[1§], Anna Mrukwa[1], Aleksandra Suwalska[1], Agata Durawa[3], Malgorzata Jelitto-Górska[3], Katarzyna Dziadziuszko[3], Edyta Szurowska[3], Pawel Bożek[4], Michal Marczyk[1], Witold Rzyman[5], and Joanna Polanska[1*]

[1] Department of Data Science and Engineering, Silesian University of Technology, Gliwice, Poland
[2] Department of Computer Graphics, Vision and Digital Systems, Silesian University of Technology, Gliwice, Poland
[3] 2nd Department of Radiology, Medical University of Gdansk, Gdansk, Poland
[4] Department of Radiology and Radiodiagnostics, Medical University of Silesia, Katowice, Poland
[5] Department of Thoracic Surgery, Medical University of Gdańsk, Gdańsk, Poland

[§] These authors contributed equally
* Correspondence: joanna.polanska@polsl.pl

## Abstract

Segmentation of the bronchovascular bundle within the lung parenchyma is a key step for the proper analysis and planning of many pulmonary diseases. It might also be considered the preprocessing step when the goal is to segment the nodules from the lung parenchyma. We propose a segmentation pipeline for the bronchovascular bundle based on the Computed Tomography images, returning either binary or labelled masks of vessels and bronchi situated in the lung parenchyma. The method consists of two modules, modeling of the bronchial tree and vessels. The core revolves around a similar pipeline, the determination of the initial perimeter by the GMM method, skeletonization, and hierarchical analysis of the created graph. We tested our method on both low-dose CT and standard-dose CT, with various pathologies, reconstructed with various slice thicknesses, and acquired from various machines. We conclude that the method is invariant with respect to the origin and parameters of the CT series. Our pipeline is best suited for studies with healthy patients, patients with lung nodules, and patients with emphysema.

## Introduction

Computed tomography (CT) is a well-recognised imaging technique widely used to support the diagnosis of pneumonia. Most commonly bronchiectasis (McGuinness et al., 1993), chronic obstructive pulmonary disease (COPD) (Ley-Zaporozhan et al., 2008), pulmonary mucormycosis (Thurlbeck & Muller, 1994), sarcoidosis (Criado et al., 2010), or emphysema (Thurlbeck & Muller, 1994). The diagnosis follows a standard procedure, in which a symptomatic patient undergoes a CT imaging screening, which the radiologist then evaluates. Expert searches for disease-specific radiological patterns. The duration of waiting time for diagnosis varies between hospitals around the world, depending on the current diagnostic load and the shortages of radiological personnel (Rimmer, 2017). However, the advent of using machine learning (ML) in radiological image analysis has brought new trends, perspectives, and possibilities to medical imaging diagnostics (Z. Zhang & Sejdić, 2019). In the case of thoracic imaging (Webb & Higgins, 2006), there exists a broad spectrum of Computer-Aided Detection (CAD) (Castellino, 2005) systems aimed at the automatic detection of common pathologies visualised by various medical imaging modalities (Retson et al., 2019). In the case of the diseases mentioned above, the focus of diagnostic evaluation lies in the appearance of the bronchovascular bundle. The automated diagnosis approach is to first segment the bronchovascular bundle and apply machine learning algorithms to learn and predict from the segmented bundle. There are also indirect benefits of bronchovascular bundle segmentation, where the bundle itself is not a source of disease but is affected by the artefacts present in the lung tissue. Such as, for example, lung nodules attached to the bronchovascular bundle. Due to connectivity, they are harder to find, either by radiologists, or CAD systems.

In the literature, researchers divide the bronchovascular bundle modelling task into pulmonary vessel segmentation and bronchi segmentation. Various approaches are used, emphasizing geometric dependencies, using machine learning and deep learning. A frequently used geometric property is the tubularity of the vessels and bronchi. Approaches using the eigenvalues of the Hessian matrix are based on this assumption and are used to characterize the model and enhance its shape (Krissian et al., 2000; Shikata et al., 2004; Sonka et al., 2009; C. Zhang et al., 2019). As the structure of the pulmonary vessels stands out strongly against the lung parenchyma, there are methods that exploit this difference by thresholding or based on volume gradients. The latter include methods using level-set segmentation group algorithms (Chen & Cohen, 2016; Cheng et al., 2015; Krissian et al., 2000; Vasilevskiy & Siddiqi, 2002). Pulmonary vessels and bronchi are fully connected structures, which is a common property in most approaches, where some of them are built on it (Kaftan et al., 2008; C. Zhang et al., 2019). More recent solutions to this problem are dominated by deep learning. They use CNNs for this (Gu et al., 2019; Nardelli et al., 2018; Zhai et al., 2019) of which the most common in the literature are different variants of U-Net (Pang et al., 2023; Ronneberger et al., 2015a; M. Wang et al., 2023; R. Wu et al., 2023) with the emphasis on its variants created using Transformer blocks (Vaswani et al., n.d.; Y. Wu et al., 2023 ).

Although many approaches have been created to solve this task, most of them are built to work with healthy lungs or are dedicated to CT studies with contrast medium. Also, little to no code or ready-to-run models can be found regarding this task, with very few public databases available to train models. Therefore, we created a processing pipeline based on CT modality, creating a detailed mask of the bronchovascular bundle with the premise of differentiating it from small abnormal structures. Bronchovascular modelling may be used as a module to be integrated by the Computer Aided Detection (CAD) system for automated diagnostics of common pulmonary diseases, or as a preliminary step in more complex procedures, for example, lung nodule detection.

# Data

In this work, we utilized a dataset gathered during the Pilot Pomeranian Lung Cancer Screening Program (Książek et al., 2009) trial in 2009-2011. We used the data to develop the resulting bronchovascular bundle model and to train the regression model. The dataset consists of 2002 patients, who have undertaken low-dose CT screening (Naidich et al., 1990) in an early lung cancer detection program. The patients were between 50 and 75 years old and had been or had smoked for at least 20 years. The main objective of the program was to detect cases of lung cancer, which could be treated with surgical operation. Patients did not present any symptoms of the disease during the trial and were considered healthy on the date of undergoing the screening.

Furthermore, we tested our algorithm on the subset of the COPDGene dataset (Regan et al., 2010; Washko et al., 2010). Which is constructed from the patient suffering from Chronic Obstructive Pulmonary Disease (COPD). Characterized by the emphysema changes visible on the lung parenchyma.

| *Dataset* | *Manufacturer* | *Slice Thickness [mm]* | *Number of series* |
|---|---|---|---|
| *Pomeranian* | GE MEDICAL SYSTEMS | 2.5 | 43 |
| | | 1.25 | 145 |
| | | 1.2 | 4 |
| | | 0.625 | 1 |
| | SIEMENS | 1.25 | 2329 |
| | | 0.625 | 1 |
| | | Unknown | 11 |
| | Unknown | 1.25 | 545 |
| | | Unknown | 21 |
| *COPDGene* | GE MEDICAL SYSTEMS | 1.25 | 7 |
| | | 0.625 | 130 |
| | Philips | 0.9 | 2 |
| | SIEMENS | 0.75 | 134 |

Table 1: Summary table showing the count of series by manufacturer and slice thickness from the Pomeranian (used for algorithms development and regression model training) and COPDGene (used for algorithm evaluation and regression model testing).

# Methods

The algorithm consists of a series of steps, each of which utilizes a unique function in the pipeline. Starting from lung segmentation followed by Gaussian mixtures modeling, preprocessing, skeletonization, finding the main direction, and finished by growing. The final output of the algorithm is a hierarchical tree of the bronchovascular bundle, where each branch forms a parent-child relationship. To fill the airways, we used a mask acquired from the GMM which served as the initial scaffolding for further processing. We skeletonized and converted the initial mask to the graph. We iterated over the nodes and applied the morphologically controlled level set segmentation. The final algorithm is capable of modeling the bronchovascular bundle in the presence of lung nodules and emphysema, but has trouble when dealing with high-attenuation patterns in the lung parenchyma. To prevent misinformation, we developed a linear model which estimates the volume of the bronchovascular tree based on the lung volume. By comparing estimate with rough values, we are able to raise a warning in the event of over-segmentation or under-segmentation. Each step is a separate algorithm, which works in a cascade, where the input of the next step is the output of the previous step. The entire pipeline is shown in Figure 1.

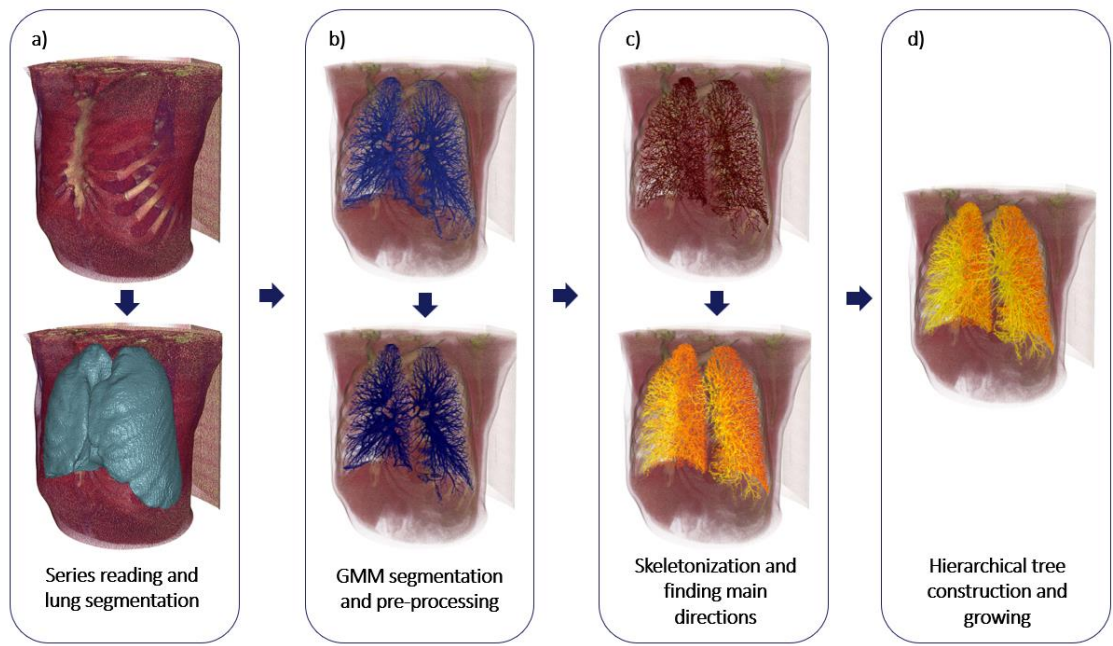

Figure 1: Processing pipeline for bronchovascular bundle modeling.

## Lung tissue segmentation

We used a Convolutional Neural Network (Albawi et al., 2018) (Yamashita et al., 2018) to segment the lung tissue from the CT volume. We utilized an architecture called U-Net (Ronneberger et al., 2015b) trained by (Hofmanninger et al., 2020), which takes as input a 2D image and outputs a binary mask. The U-Net architecture is dedicated to segmentation and is broadly used in the medical imaging field (Du et al., 2020). The diseases present in the authors' dataset contained cases of fissures, air pockets, as well as tumours, and effusions. The model generates good quality segmentations of both left and right lungs for healthy patients and patients who have pneumonia-like diseases, such as COVID-19.

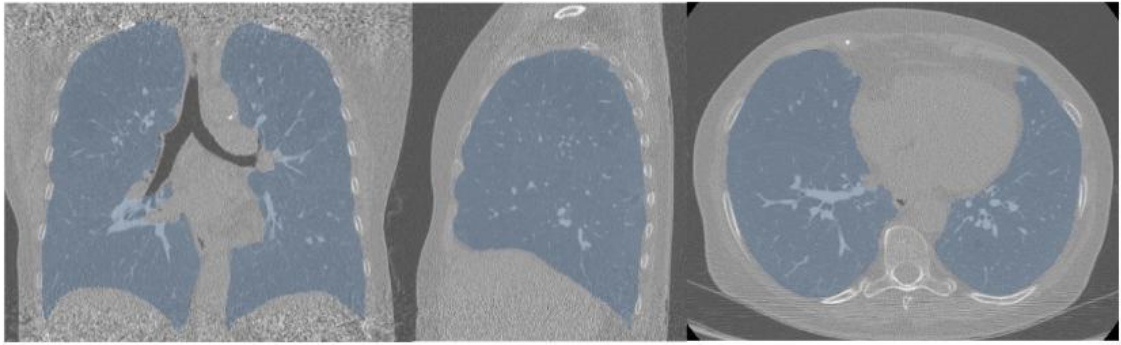

Figure 2: Lung segmentation viewed from a) coronal, b) sagittal, and c) axial planes.

Segmentation was carried out sequentially, where consecutive axial slices of the CT volume were passed as input to the 2D network. The goal of the neural network was to associate a class for each pixel in the image with either the background or lung tissue class. The input is transformed by the contractive and expansive U-Net paths and a binary segmentation map of the lungs and background is generated.

### *Trachea Segmentation*

We segmented the mediastinum by iterating over the axial plan while creating the convex hull. From the masked mediastinum, we segmented the trachea. We thresholded the acquired mediastinum by setting the cut-off thresholds in its minimum value and the third of its minimum (which is approximately -1024 and -341). The acquired mask contains the trachea and other structures. To acquire the trachea, we used connected components labelling which resulted in the set of unconnected masks $M$. To distinguish the trachea from other structures, we exploit two properties: it is centrally located and it is one of the largest air-filled elements of the mediastinum. For each mask, we calculate its area, storing it in vector $A$ and the minimal distance from the centre of the image, storing it in vector $D$. We normalize the values in the $A$ and $D$ vectors and create the vector $D_r$ using equation $D_r = |1 - D|$. Then we calculated the $I$ vector as the weighted average of vectors $A$ and $D_r$ with consecutive weights 2 and 1. The maximum value of the vector $I$ indicated the component which was the trachea mask.

### *Preprocessing*

The trachea mask was enlarged using the dilation operation with the 3D box structuring element of size (20, 20, 20). We added the trachea to the lung mask,, creating a binary mask then iterated over the slices in the axial plane and eroded them with the 2D box structuring element of (1, 1, 1). In this way, we removed noise in the form of over-segmentations of the lung mask (e.g., small fragments of ribs) while protecting the main bronchus and adjacence vessels from degeneration.

*Gaussian Mixture Modeling Quantization*

We used Gaussian Mixture Modelling (Reynolds, 2009) to cluster the voxel intensities in the segmented lung tissue area. We used three components to model the distribution of the voxel intensities due to the nature of density variations in the lung tissue. The class assignment for the given model is performed by the maximum likelihood function of the Gaussian Mixture components. In other words, we label each voxel by the class of its component, to which the voxel most likely belongs.

The first component aggregated voxels of the lowest intensity, including the airways segmented together with the lung tissue, as low-density well as lung artifacts, such as emphysema. The second component clustered the overall lung tissue in the segmented volume. Finally, the third component grouped the voxels with the highest intensities. These voxels constituted the bronchovascular bundle as well as artifacts found in the lung tissue, such as nodules, calcifications, fibrosis, or inflammations. The step fulfills two tasks: not only does it provide masks of the object present in the lung tissue, which is often a preliminary step for classification tasks, such as radiomic-based classifiers (Prazuch et al., 2022), but it also provides a segmentation of the full bronchovascular bundle for the remaining steps in the pipeline.

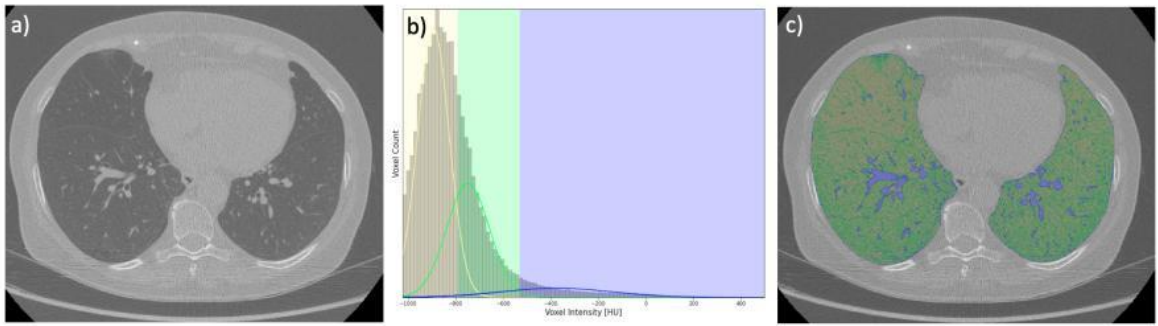

Figure 3: Quantization of the voxels of lung tissue by GMM components. a) Original slice. b) GMM quantization. c) Quantized voxels mapped onto the original CT volume.

Using the connectivity criterion, we extracted the largest objects segmented by the third GMM component. Usually, the number of objects ranges between 2 (one for each lung) and 6 (one for each lobe, or even segment), depending on the anatomical structure of the lungs related to lung fissures (Meenakshi et al., 2004). We use a knee plot to estimate the exact number of those independent objects by labeling the isolated objects with a unique class. We then draw an object voxel class count and find the knee of the plot. We treated objects whose voxel count is greater than the count of knee point as a whole bronchovascular bundle and removed the rest of the voxels by assigning a background class to them. From the segmented objects, we create a bronchovascular mask which is mapped to the space of the original volume.

## *Skeletonisation of the Bronchovascular Bundle*

We performed skeletonization (Abu-Ain et al., 2013) of the segmented bronchovascular volume mask. In order to perform it in the 3D space, we used Lee's algorithm (Lee et al., 1994). The algorithm takes as input a binary mask of the bronchovascular bundle and processes the input in a sequential manner, where in each step the thickness of the branches is reduced by 1 with additional checks on connectivity preservation in the 3D space (Cohen-Or & Kaufman, 1997). The branches of the bronchovascular bundle are transformed into a simplified model of the structural map. Each branch and node in the bundle is a set of points with 1 unit thickness. This step forms a simplified model of the bronchovascular bundle, which is a preliminary step before forming a graph of the bundle.

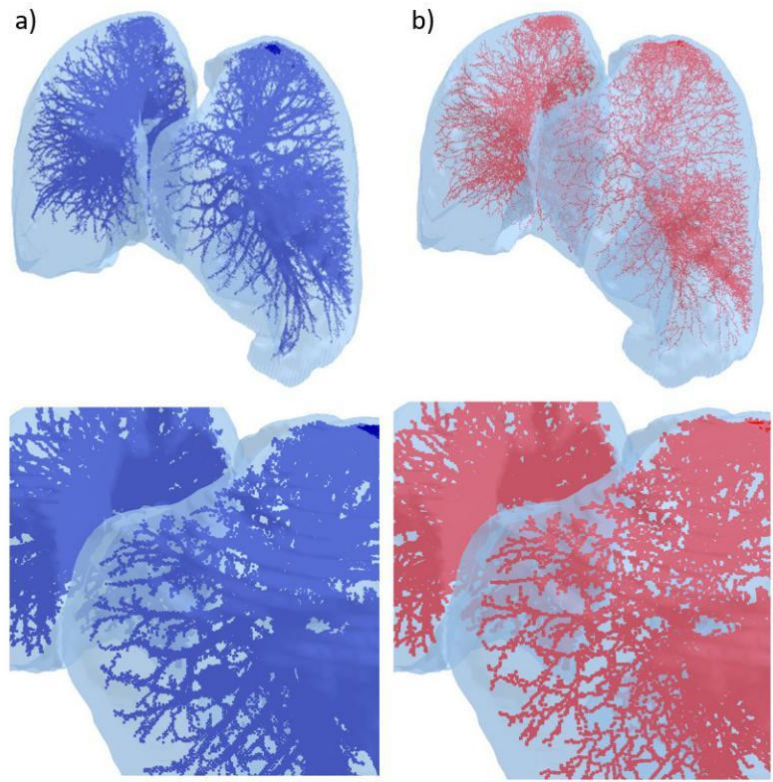

Figure 4: The result of skeletonisation of the segmentation of the bronchovascular bundle. a) Original segmentation of the bronchovascular bundle. b) Skeleton created from the segmentation.

## **Graph forming**

We used the 'sknw' package provided in the ImagePy (A. Wang et al., 2018) library to create graphs for the skeletonised bronchovascular bundle. The package located the nodes in the skeleton using the neighborhood criterion. If a given voxel was surrounded by more than two skeleton pixels, it was recognized as a node voxel class; otherwise, it was treated as an edge class. The algorithm performed the neighborhood check for all skeleton voxels creating the graph model. Each node stored information about the node center and the coordinates of the pixels that were used to form a node. Each edge of the graph stored information about the distance between connected nodes and a list of voxel coordinates that formed the edge. Finally, it produced the object, which kept track of all the nodes and edges detected in the structure.

Each branch was also described by a set of two unique identifiers. These identifiers refer to the starting and end nodes. For simplicity, for each identifier, a simple integer-based identifier was added for each branch.

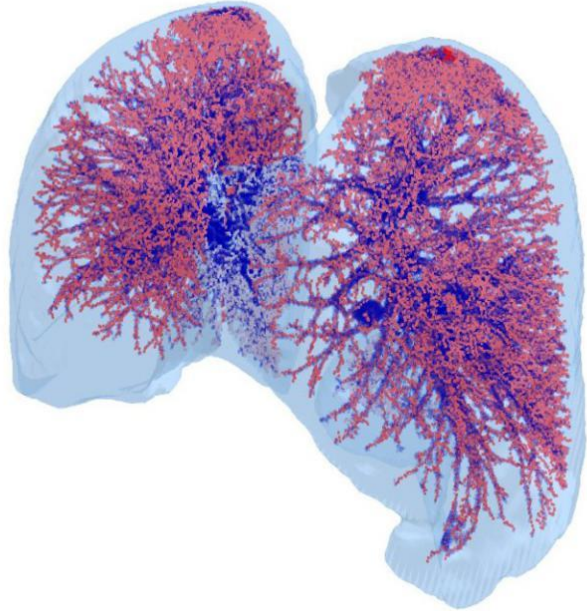

Figure 5: Graph of the skeletonized bundle mapped to the original segmentation of the bronchovascular bundle.

## *The main direction searching*

For each skeleton branch, the main direction was found by calculating the unit vector $\vec{U}$ between the starting and end node coordinates. We defined base vectors $\vec{N}$ to be normal vectors of axial, sagittal, and coronal planes. Then we used cosine similarity to find the most similar base vector $\vec{V}$ for that direction (**Eq. 1**). Treating the two remaining orthogonal base vectors as growing directions for our iterative branch label assignment algorithm.

$$\vec{V} = argmin\left(\frac{\vec{U} \cdot \vec{N}}{\|\vec{U}\|\|\vec{N}\|}\right) \qquad \vec{N} \in \{\vec{N}_{ax}, \vec{N}_{sag}, \vec{N}_{cor}\} \qquad (1)$$

## *Hierarchical Tree Construction*

After finding the major direction for all the branches, we grew each skeleton branch by voxel unit in each orthogonal direction. We found 8 directions for a branch in quantized voxel space. After performing the growing step, we used a boolean AND operation between the grown region and the original GMM bundle mask. The result of the operation is treated as an initial step for the next iteration of the procedure, where each branch is grown again in the directions of orthogonal vectors. We repeat the process until each voxel from the GMM mask is assigned a branch-label class.

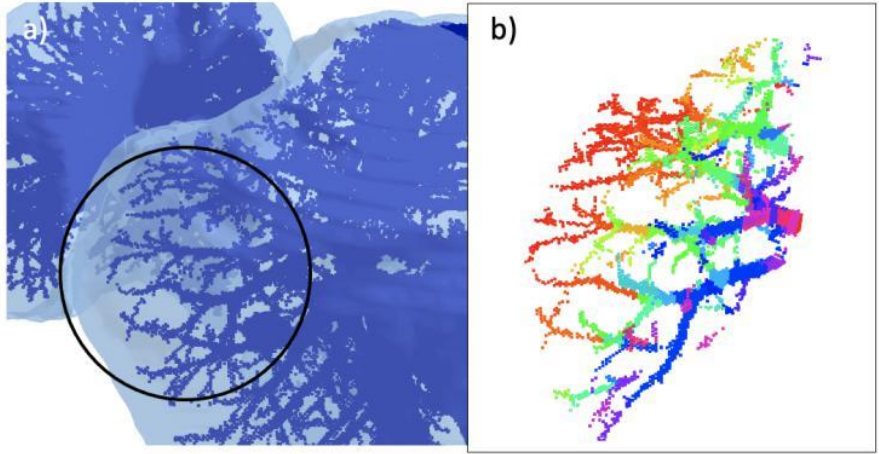

Figure 6: Labeling of the branches in the bundle. a) Sub-volume of the original bundle mask. b) Labeled branches marked by different colors.

## Bronchi modeling

We used the inverse mask of the last GMM component as an initial mask for the construction of the bronchi model. We eroded the initial mask with a 3D ball structuring element of radius equal to 1 voxel. The initial mask and trachea were skeletonized and transformed into two graphs. In the trachea graph, we found the node with the highest index in the axial plane and looked for the closest node to it in the initial mask graph marking it as the starting node.

First, we defined the speed as the gradient magnitude (**Eq. 2**) mapped with the bounded reciprocal $1/(1+x)$. Thus, we created a speed image with the largest value equal to one in areas away from the edge and voxel values converging toward zero the closer they are to the edge.

$$\parallel \nabla I(x,y,z) \parallel = \sqrt{=\left(\frac{\partial x}{\partial I}\right)^2 + \left(\frac{\partial y}{\partial I}\right)^2 + \left(\frac{\partial z}{\partial I}\right)^2} \qquad (2)$$

Then from the starting node, we iteratively walked through the neighboring nodes and used the fast-marching algorithm (Sethian, 1999) with the seed point in the node's center. We masked the speed image with the acquired segmentation, calculating the sprawl distance by summing the values. We compare the sprawl distance of the current node with its neighbors. If the sprawl distance of the current node is greater than the sprawl distance from the previous node, we suspected a leak in the initial mask.

To counter the leak we performed controlled 3D erosion with the ball structuring element and radius size of 1 voxel on the binarized result of the fast marching algorithm. The erosion is performed at a maximum of 5 times until the number of connected components rises or stays the same. We counted erosions that were number of the performed and then applied the same amount of dilation operations. This way on the resulting mask, the part of the bronchi and the result of leakage should be separated. We find the mask of separation and subtract it from the speed image. Then the fast marching algorithm is used on the same node but with a modified speed image. If the sprawl distance is still bigger than the sprawl distance of the previous node, the current node is removed from the graph and the speed image at that point is blocked. Final segmentation is a binarized sum of all results from the fast-marching algorithm.

### *Volume Calculation and Estimation*

We calculated the volume of the resulting bronchovascular bunlde and lung segmentation using the SimpleITK package that gives us the exact value of the volumes. Using the Pomeranian dataset volume values, we constructed the linear regression model to estimate the bronchovascular bundle volume based on the exact lung volume. Then we calculated the confidence intervals as well as the Chauvenet criterion for the predicted volumes. When the segmentation of the volume calculated from the bronchovascular tree is not in the expected range, the alternative value of the regression model can be proposed as the appropriate volume.

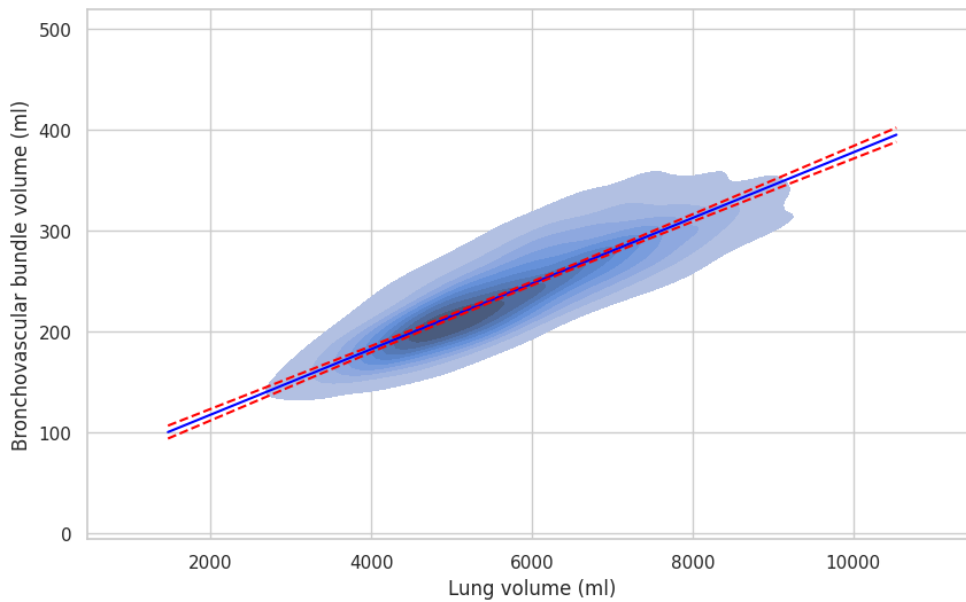

Figure 7: Linear regression models (blue line) and confidence intervals (red dotted line) for bronchovascular bundle volume calculation with isolines of the training dataset (Pomeranian dataset) points density.

## Results

We generate a set of objects compatible with the DICOM format as the resulting output bundle. The resulting lung segmentation mask provided by (Hofmanninger et al., 2020) is converted to the DICOM format under the same study instance UUID and the bronchovascular bundle is labeled according to the hierarchical tree or binary depending on the function parameters. We tested the method on CT images of healthy patients (**Figure 8a**), patients with lung cancer (**Figure 8b**), and patients with COPD (**Figure 8c**).

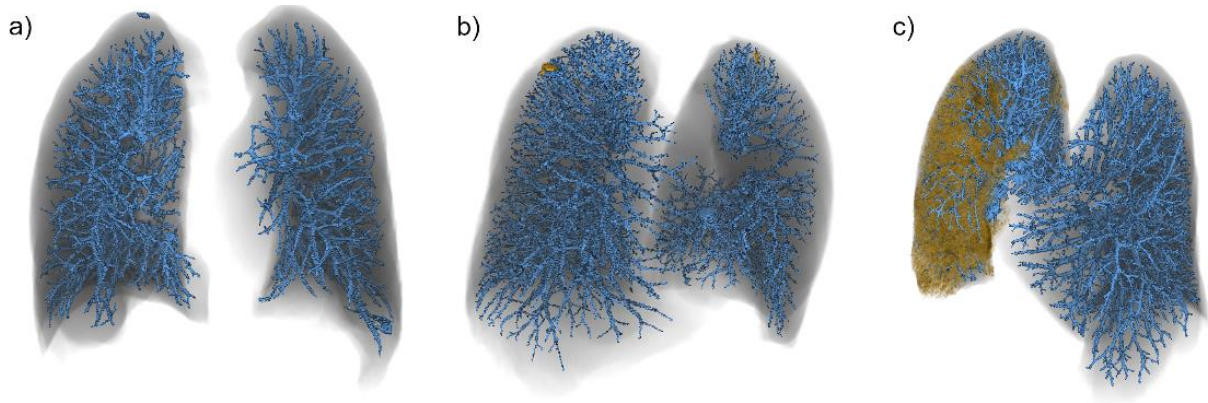

Figure 8: Example binarized bronchovascular bundle segmentation shown in the example of a) healthy lungs, b) lungs with lung nodules, and c) lungs affected by emphysema.

The algorithm performed well in the mentioned examples. In **Figure 7** the resulting bronchovascular bundle model is shown in the blue, and abnormalities in lung parenchyma are shown in yellow. The constructed linear model was tested on the COPD dataset (**Figure 9**), which showed to be approximating the segmented volumes in a satisfactory manner. The volume predicted from the regression model was mostly higher than the one resulting from the bundle segmentation (**Figure 10**) due to the higher resolution of the test dataset.

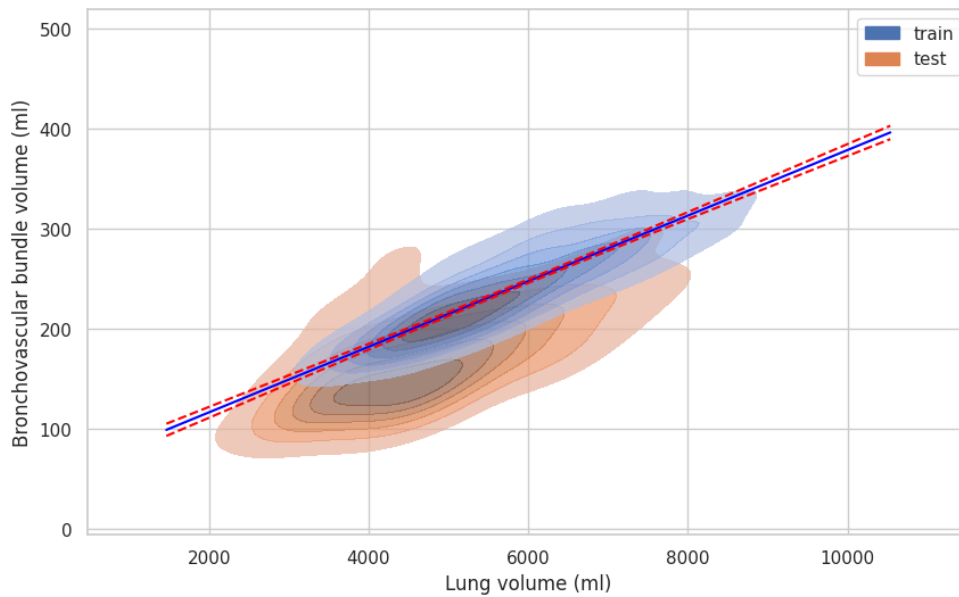

Figure 9: Results of bronchovascular volume calculation for lungs affected by emphysema in relation to lung volume with regression line indicating prediction of volume of the segmented bundle.

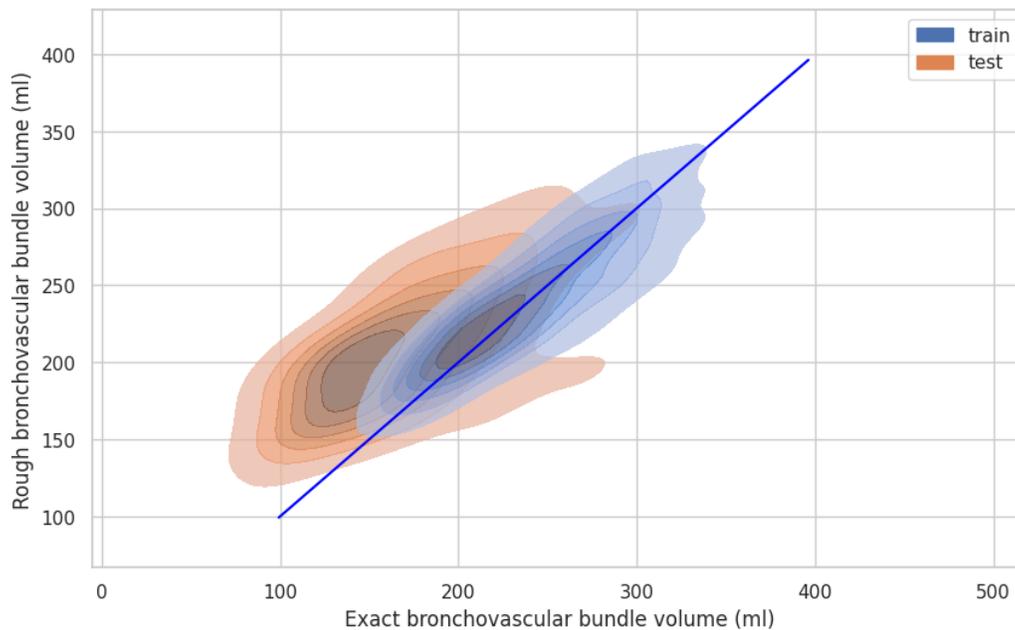

Figure 10: Results of bronchovascular volume calculation using the regression model for lungs affected by emphysema in relation to the bronchovascular volume of the segmented bundle.

# Discussion

## *Nodule adjacenc to the bundle*

There are cases where artifacts of lung tissue are present and are connected to the bronchovascular bundle. This case is depicted in **Figure 11**. During branch and node modeling, such nodules on rare occasions are falsely treated as a part of the ordinary structure of the bundle, as the algorithm models the nodule as a branch with a starting and end node. For such cases, additional post-processing based on the shape and monotonicity of the slices of the branches is needed. One way to approach it is to verify the monotonicity of the thickness of the branch compared to the parent and child branches. The nodule adjacent to the affected branch would likely not follow the contracting thickness behavior as the tree is traversed from the root node to the child nodes.

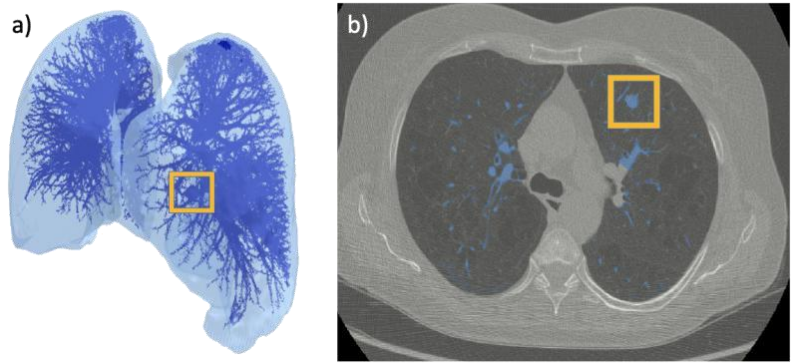

**Figure 11**: a) Adjacent nodule to the bronchovascular bundle in our generated model. b) The same nodule is shown on the CT slice in the open source viewer.

## High attenuation patterns

Diseases such as pneumonia, sarcoidosis, tuberculosis, or COVID-19 are characterized by high attenuation patterns such as consolidations or ground glass opacities present in lung parenchyma. For such diseases, the proposed method does not model the bronchovascular bundle well. If there are such cases in the processed set, our system will detect them and notify the user. To estimate the volume of the bronchovascular bundle in such a case, we suggest using the volume estimator instead of the volume calculated from the modeled bronchovascular bundle.

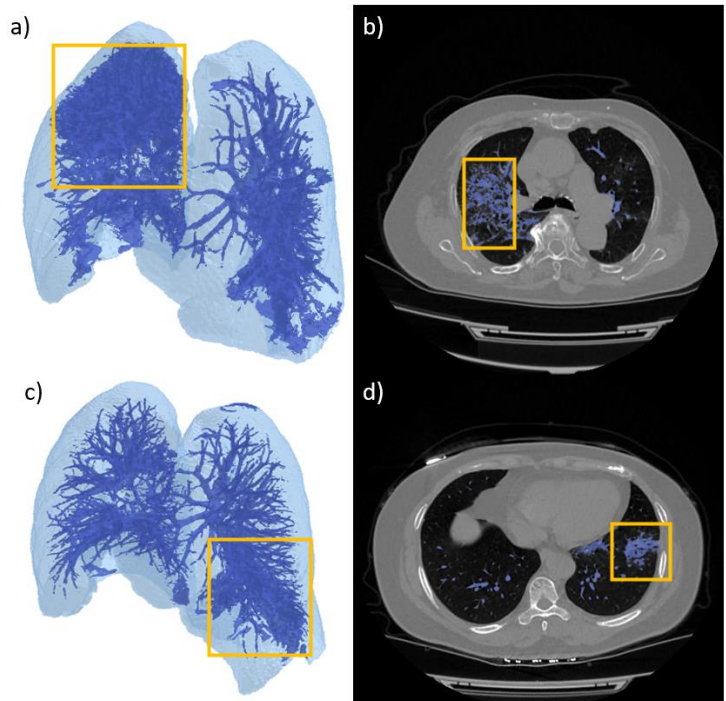

Figure 12: Example modeling results acquired from series with lung parenchyma afflicted by high-attenuation patterns a,b) diffuse consolidations highly spread through the lungs, c,d) consolidations concentrated in one lung area.

The results of the bronchovascular modeling from both of the series presented in **Figure 12** were verified by the regression model and results were overlayed on the regression line (**Figure 13**).

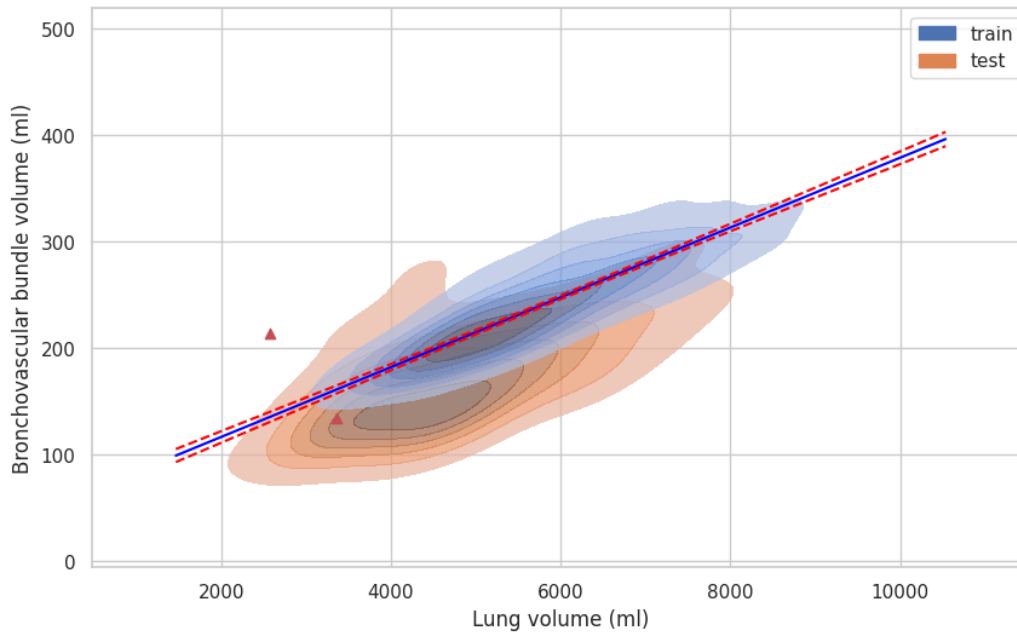

Figure 13: The predictions of the resulting model (red triangles) from the series presented in Figure 9. The results of the series presented in Figure 9ab are located farthest from the regression line. The results for the series presented in Figure 9cd are located in the orange isolines, much closer to the regression line.

The results for the series located in **Figure 12ab** would be marked as inappropriate by the regression model. Its bronchovascular modelling result is much bigger than the estimated value. As for the results for the series presented in **Figure 12cd**, it would depend on the assumed threshold for the 2D Gaussian isolines. In the current example, it would also be regarded as suspicious.

## Voxel quantization approaches

We designed a 3-component-based GMM quantization approach to cluster the bronchovascular bundle voxels together. However, different approaches for voxel quantization could be explored. For example, in our work for COVID-19 inflammation clustering, we used a 6-component local Gaussian Mixture Modeling called MimSeg to differentiate between different types of inflammation in the lung parenchyma. For patients with COVID-19 pneumonia, there are cases of ground-glass opacity-type inflammations as well as consolidations. Those inflammations are characterized by varying densities of the affected regions, and thus, can be distinguished by the GMM-based thresholding performed by MimSeg. We show the voxel intensity thresholds calculated by MimSeg in **Figure 14** and discuss the algorithm in (Binczyk et al., 2017).

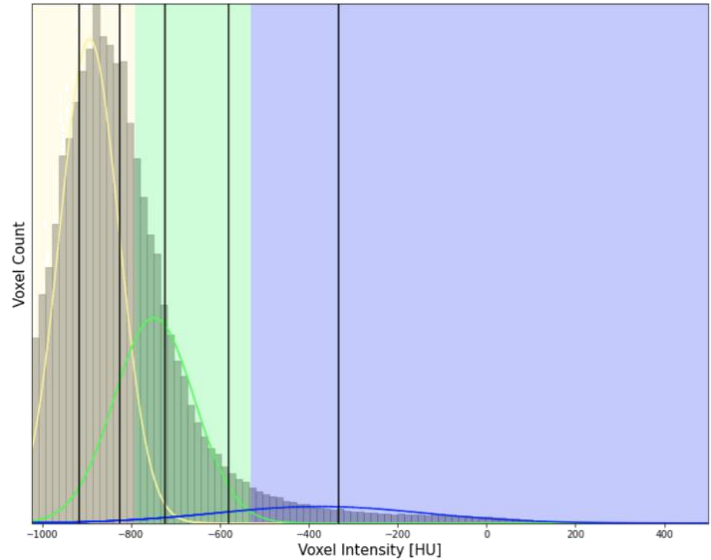

Figure 14: MimSeg thresholds shown on the original GMM quantization histogram. MimSeg creates 5 thresholds, which divide the voxel intensity domain into 6 sections. In this case, the two highest intensity thresholds could be used to group the bronchovascular bundle voxels.

## Conclusions

We prepared a bronchovascular bundle modeling algorithm, which creates a hierarchical structure of branches and nodes reflecting the structure of the bronchovascular bundle. This bundle can be used to analyse the relationship and radiological characteristics of the airways, as well as the blood veins comprising the model. Such a solution may be adopted in CAD systems supporting the work of radiologists or could automate the diagnostic procedures of common bronchovascular diseases diagnosed using the Computed Tomography imaging technique. In addition, the algorithm can be used in detecting lung nodules adjacent to the bronchovascular bundle, as in the case shown in **Figure 8**.

This algorithm can be easily integrated into radiological consoles because it is compatible with common medical imaging data formats. We open source the code for our algorithm on https://github.com/ZAEDPolSl/BRONCO.

## Acknowledgements

This research was funded by the National Science Center, Poland, Grant 2017/27/B/NZ7/01833. Clinical material was collected in the frame of the project MOLTEST-BIS (DZP/PBS3/247184/2014). Additionally, AS and WP are holders of European Union scholarship through the European Social Fund, grant POWR.03.05.00-00-Z305.